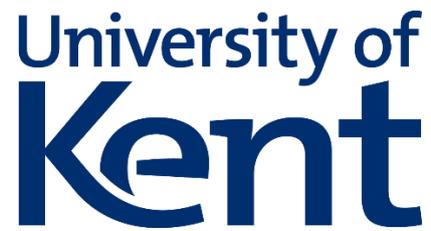

Evasion-Resilient Detection of DNS-over-HTTPS Data Exfiltration: A Practical Evaluation

and Toolkit

By:

Adam Elaoumari

ae469@kent.ac.uk

School of Computing

University of Kent

MSc Cyber Security

2025

9751 words

Supervised by

Dr Darren Hurley-Smith

# UNIVERSITY OF KENT

## ACCESS TO A MASTER'S DEGREE OR POSTGRADUATE DIPLOMA DISSERTATION

In accordance with the Regulations, I hereby confirm that I shall permit general access to my dissertation at the discretion of the University Librarian. I agree that copies of my dissertation may be made for Libraries and research workers on the understanding that no publication in any form is made of the contents without my permission.

**Notes for Candidates:** by submitting your dissertation, you agree with the following:

1   Where the examiners consider the dissertation to be of distinction standard, one copy may be deposited in the University Library and/or uploaded into Moodle as an example of good dissertation for future students.

2   If a copy is sent to the Library, it becomes the property of the University Library. The copyright in its contents remains with the candidate. A duplicated sheet is pasted into the front of every thesis or dissertation deposited in the Library. The wording on the sheet is:

> *"I undertake not to use any matter contained in this thesis for publication in any form without the prior knowledge of the author."*

Every reader of the dissertation must sign and date this sheet.

3   The University has the right to publish the title of the dissertation and the abstract and to authorise others to do so.

..................................................................................................................................

**SIGNATURE**                                                                                          **DATE 30/08/2025**

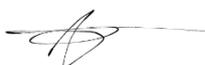

...........................................................................................................................
**Adam ELAOUMARI**

---

### CERTIFICATE ON SUBMISSION OF DISSERTATION

I certify that:

1. I have read the University Degree Regulations under which this submission is made;

2. In so far as the dissertation involves any collaborative research, the extent of this collaboration has been clearly indicated; and that any material which has been previously presented and accepted for the award of an academic qualification at this University or elsewhere has also been clearly identified in the dissertation.

...........................................................................................................................
**SIGNATURE**                                                    **DATE 30/08/2025**

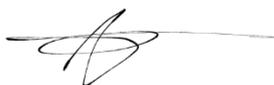

This form should be completed and included in the submission of your dissertation.

# List of figures

# List of tables

# Abstract

The purpose of this project is to assess how well defenders can detect DNS-over-HTTPS (DoH) file exfiltration, and which evasion strategies can be used by attackers. While providing a reproducible toolkit to generate, intercept and analyze DoH exfiltration, and comparing Machine Learning vs threshold-based detection under adversarial scenarios.

The originality of this project is the introduction of an end-to-end, containerized pipeline that generates configurable file exfiltration over DoH using several parameters (e.g., chunking, encoding, padding, resolver rotation). It allows for file reconstruction at the resolver side, while extracting flow-level features using a fork of DoHLyzer. The pipeline contains a prediction side, which allows the training of machine learning models based on public labelled datasets and then evaluates them side-by-side with threshold-based detection methods against malicious and evasive DNS-Over-HTTPS traffic.

We train Random Forest, Gradient Boosting and Logistic Regression classifiers on a public DoH dataset and benchmark them against evasive DoH exfiltration scenarios. The toolkit orchestrates traffic generation, file capture, feature extraction, model training and analysis. The toolkit is then encapsulated into several Docker containers for easy setup and full reproducibility regardless of the platform it is run on.

Future research regarding this project is directed at validating the results on mixed enterprise traffic, extending the protocol coverage to HTTP/3/QUIC request, adding a benign traffic generation, and working on real-time traffic evaluation. A key objective is to quantify when stealth constraints make DoH exfiltration uneconomical and unworthy for the attacker.

# Table of contents

# 1. Introduction

## 1.1 Research Context

The Domain Name System (DNS) underpins how users and applications locate services on the Internet.

The DNS protocol have a clear define use cases (defined in the RFCs), despite being capable of more. Attackers took advantage of the capabilities of DNS, and so DNS traffic has long been abused for illegal activities and used as a covert communication channel for command-and-control server and data exfiltration because it is a ubiquitous protocol that is often loosely monitored [1].

Designed in 1983 and first standardized by RFC 882/883 [2], the DNS protocol was overhauled in 1987 by RFC 1034/1035 which still constitutes its basis [3]. It has since been extended by EDNS(0) which was first proposed in RFC 2671 and later obsoleted by RFC 6891 [4] and secured by DNSSEC [5]. Over the years, a security and privacy issue has arisen regarding the DNS protocol. The latter is always using clear traffic, it could be used by government or malicious entities to analyse and modify DNS responses, allowing governments to block access to websites, in the context of censorship for example.

This is how DNS over HTTPS emerged as a solution to these security and privacy issues.

DNS-over-HTTPS (referred to as DoH in this dissertation), standardized in 2018 with RFC 8484 [6] encapsulates DNS queries inside HTTPS streams over port 443, thus blending them within ordinary encrypted web traffic. This design that acts on the application layer of the OSI model prevents observers from directly inspecting DNS lookups, whilst protecting the users from man-in-the-middle attacks.

Following its publication, DoH rapidly gained operating-system support, as Microsoft integrated it in Windows 10 (build 20185) and is supported by default in Windows 11 [7], this represents 96.8% of Windows users [8]. Google deployed it in Android 11 released in September 2020 [9] and added Chrome support for DoH in version 83 whilst Apple enabled it in iOS 14 and macOS 11. Adoption however has been uneven, some browsers such as Firefox activate DoH by default in certain region in partnership with Cloudflare [10], while Chrome uses DoH by default since May 2020 if the user's default DNS server already supports this protocol in the first place [11].

This section is partially adapted from a previous coursework submitted at the University of Kent, for the Privacy module, that showcased DoH as Privacy-Enhancing-Tool for everyday users. In contrast, the present dissertation aims to explore the darker side of the protocol, how its privacy features can be exploited by attackers for data exfiltration, and what security challenges this poses for defenders [41].

DoH's promise (making DNS look like ordinary HTTPS on port 443) also creates a security paradox: defenders lose straightforward DNS telemetry (domains, RR types, timing patterns) that traditionally powered intrusion detection, DLP controls, and incident response. The resulting policy debate weighs privacy benefits against potential blind spots for enterprises and ISPs. It is worth noting that DoH can be implemented at multiple layers, either at application-level (e.g. browsers), OS level resolver (e.g. Windows), gateway/router or enterprise/public resolver which complicates monitoring, policy enforcement and incident response.

## 1.2 Problem

When DNS traffic is encrypted and multiplexed with web traffic, adversaries can exfiltrate data over DoH while blending into benign HTTPS patterns. Prior research showed effective detection of classic DNS covert channels and low-throughput exfiltration in plaintext DNS [13], but DoH complicates this by obscuring domain names and observable features. Public datasets such as CIRA-CIC-DoHBrw-2020 [14] and tools like DoHLyzer [15] helped establish DoH/non-DoH and benign/malicious DoH classification pipelines, yet these datasets are hard to reproduce in real life, as setting up a test lab for DoH is hard and time consuming. Also, multiple proposed detections methods report high accuracy on benchmark datasets but leave open questions about robustness to adaptive attackers who shape timing, size, padding, and resolver behaviour to evade detectors [13, 16]. This dissertation addresses that gap by studying evasion-aware detection, and by releasing an open toolkit that industrialize both the attack using configurable exfiltration profiles and the defence with built-in feature extraction, heuristic, and misclassification tracing, using both ML models and threshold-based models such as DoHxP [17].

## 1.3 Contributions

### 1.3.1 Theoretical Contributions

This dissertation makes theoretical contributions to the understanding of DoH security, by offering a critical analysis of existing detection approaches and highlighting their limits in the face of evasion techniques. It makes it possible to identify the levers available to attackers to manipulate traffic characteristics, such as the frequency of packets, the size of chunks transmitted, the volume, the size of payloads or even the modification of the IP stack (source IP) to alter the generation of flows by detection tools. However, the latter is not practicable, as this would impact TLS handshake and prevent the handshake to occur properly. This dissertation proposes the implementation of a threshold-based detection system, DoHxP [17], which although simple, offers a structured framework to evaluate the robustness of detection mechanisms in the face of escape behaviour. Finally, it enriches academic literature by highlighting the gap between laboratory evaluations, often based on balanced datasets, and the complexity of real traffic where malicious activities represent a marginal fraction of the traffic analysed.

### 1.3.2 Practical Contributions

On a practical level, this dissertation proposes and evaluates a reproducible experimental framework for the study of data exfiltration via DoH as well as its detection. The work sets up a complete infrastructure, including a custom DoH server, an HTTPS reverse-proxy, a controlled domain as well as exfiltration and detection tools adapted to this infrastructure, all allowing the generation of realistic traffic as part of the evaluation of a model, or benchmarks of evasion techniques. It also extends existing tools, notably a fork of DoHLyzer [15] used during the creation of the CIRA-CIC-DoHBrw-2020 dataset [14] that this dissertation uses. This dissertation also introduces automation scripts, enabling the user to do large-scale tests, with a structured export of the results in JSON, allowing for easier visualization and analysis of the results. By using this tool to conduct experiments on both public datasets and samples of these, the impact of different evasion strategies on detection accuracy and results is highlighted. This work results in the provision of a modular toolkit, based on Docker for greater reproducibility and rapid deployment, that researchers and analysts can reuse to test detection approaches in adverse conditions. The toolkit is open source, allowing others to interrogate, validate and contribute to the tool capabilities.

## 1.4 Project Aim

The aim of this project is to design, implement and evaluate a reproducible experimental framework for detecting data exfiltration over DNS-over-HTTPS (DoH), with a particular focus on analysing how adversarial evasion strategies impact the effectiveness and robustness of existing detection approaches. In the end, the goal of this project is to not only maximise detection performance but enable researchers to develop methods that either detect exfiltration at useful rates, or force adversaries to use such low exfiltration that it is not viable anymore, making any undetected DoH exfiltration impractically slow.

## 1.5 Project Objectives

To achieve the project aim, these objectives have been defined:

1. Conduct a critical state-of-the-art review of DNS-Over-HTTPS malicious traffic detection methods and existing evasion strategies.
2. Deploy a controlled DoH environment that includes a custom DoH server, a local domain name, and DoH file exfiltration tools to generate malicious traffic.
3. Use and adapt DoHLyzer to extract flow characteristics.
4. Train different models on labelled public datasets.
5. Design a prediction module that includes:
   - Machine learning models trained on public datasets [14]
   - A heuristic method based on statistical thresholds (DoHxP [17])

   To compare their respective performances

6. Develop different evasion techniques (e.g. slow exfiltration, fragmentation) to evaluate the robustness of detection methods
7. Perform detection tests on previously trained models via public datasets
8. Consolidate all features into a modular toolkit, using Docker to ensure reproducibility and reusability

## 1.6 Dissertation Structure

This dissertation is structured into seven chapters. First, Chapter 2 will present a comprehensive literature review, offering a critical analysis of existing detection approaches for DNS-over-HTTPS, ranging from threshold-based detection techniques to machine learning methods, while also examining adversarial evasion strategies. Chapter 3 then outlines the research methodology and experimental design, describing the datasets employed as well as the infrastructure setup, including the DoH server, traffic generation tools and feature extraction process. Chapter 4 presents and interprets the experimental results across labelled baselines and adversarial scenarios. Chapter 5 is a discussion of the findings and the practical contributions of this dissertation, whilst Chapter 6 outlines avenues for future work. In the end, Chapter 7 concludes the dissertation

## 2. Literature Review

The purpose of this review is to identify techniques and problems associated with these techniques, whilst assessing the current state-of-the-art regarding DNS-Over-HTTPS file exfiltration detection.

### 2.1 File exfiltration using DNS-Over-HTTPS

Exfiltration using DNS-Over-HTTPS uses encoded data fragments inside the subdomain's labels (that are constrained by the protocol the label must be less than or equal to 63 octets and the FQDN must be less than or equal to 255 octets), which the attacker resolves using a server under his control. The server then reconstructs the payload from successive requests. In DoH, these DNS messages are transported in HTTPS exchanges using either the GET protocol typically with a parameter "?dns=" containing the DNS request or the POST protocol with an "application/dns-message" body, merging with encrypted web traffic on port 443 [6]. This greatly reduces network observability and bypasses filtering policies based on clear DNS, as the traditional filtering methods rely on reading the DNS queries, which is not possible in this case. The usual techniques combine chunking/encoding, low-and-slow exfiltration to dilute the traffic with legitimate traffic. Beyond file exfiltration, malicious actors can also operate command-and-control (C2) over DoH as a covert channel, further complicating attribution and defences. Documented campaigns (e.g., ChamelDoH [18]) confirm this operational use of DoH for C2 and exfiltration. As noted by APNIC [19], DoH increasingly blends into ordinary HTTPS and is often enabled at the application layer ($7^{th}$ layer of the OSI model), which further limits network-layer inspection at enterprise perimeters. In practice, campaigns like ChamelDoH have leveraged public DoH endpoints to route C2 over port 443 and sidestep DNS-layer logging [18]. Figure 1 illustrates this threat model: a compromised host (optionally via a local DoH proxy) issues encoded subdomain queries to a DoH resolver, upstream resolution reaches an attacker-controlled authoritative server that reconstructs the exfiltrated data and returns commands, all whilst being encapsulated within HTTPS.

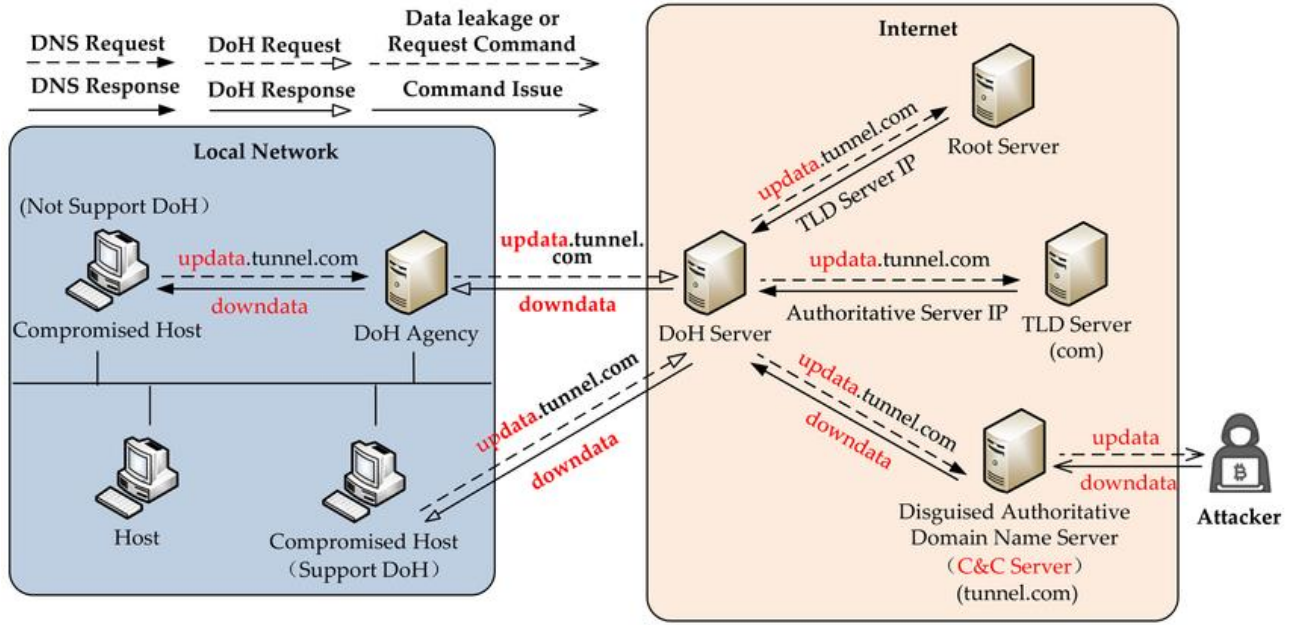

*Figure 1. Data leakage and command control process based on DoH-encrypted DNS covert channels. (Wang, Y., Shen, C., Hou, D., Xiong, X., & Li, Y. (2022). FF-MR: A DoH-encrypted DNS covert channel detection method based on feature fusion. Applied Sciences, 12(24), 12644. https://doi.org/10.3390/app122412644)*

## 2.2 Datasets and reference tools

### 2.2.1 Feature representation

We rely on the official DoHMeter flow-level CSVs provided with CIRA-CIC-DoHBrw-2020 dataset [14, 20]. Each row corresponds to a single DoH flow labelled Benign or Malicious. For modelling, we use all numeric feature from the release and keep metadata (IPs and timestamp) only for splitting and analysis. Feature families are summarized in Table 1.

*Table 1. Feature families of the CSVs files used for training*

| Family | Columns | Units | What it captures |
|---|---|---|---|
| Bytes & Rates | FlowBytesSent, FlowSentRate, FlowBytesReceived, FlowReceivedRate | Bytes, Bytes/s | Per-flow volume and average transfer rates |
| Packet-Length statistics | PacketLength (Mean, Median, Mode, Variance, StandardDeviation, CoefficientofVariation, SkewFromMedian, SkewFromMode) | Bytes, bytes$^2$, unitless | Central tendency, dispersion, asymmetry |

| | | | of packet sizes |
|---|---|---|---|
| Inter-packet-time statistics | PacketTime(Mean, Median, Mode, Variance, StandardDeviation,CoefficientofVariation, SkewFromMedian, SkewFromMode) | Seconds, seconds$^2$, unitless | Timing structure between packets |
| Request ↔ Response delta statistics | ResponseTimeTime(Mean, Median, Mode, Variance,StandardDeviation, CoefficientofVariation, SkewFromMedian,SkewFromMode) | Seconds, s$^2$, unitless | App-level RTT (Round-Trip-Time) proxy (request ↔ response deltas) |
| Metadata | SourcePort, DestinationPort, Duration | Int, float | Metadata |
| Metadata (not used for training) | SourceIP, DestinationIP | String (IPv4/IPv6) | Ground truth |

## 2.2.2 Public dataset and tools

This dissertation uses a public dataset for training and evaluating the models, specifically CIRA-CIC-DoHBrw-2020, a two-layer dataset that captures both benign DoH, malicious DoH (generated with dns2tcp, DNSCat2, Iodine) and non-DoH HTTPS. Benign traffic was generated by automated browsing of the top 10k Alexa sites with Chrome and Firefox, whilst DoH queries were sent to resolvers such as AdGuard, Cloudflare, Google and Quad9. Traffic was captured between a DoH proxy and DoH servers with tcpdump, and malicious requests were sent using a random value as transmission rate (between 100 to 1100 bits per second) [14]. The dataset is distributed into two forms: raw PCAPs and pre-computed CSVs. DoHLyzer [15] denotes the overall toolset used by the authors to parse PCAPs/sniff live DoH traffic and structure flows using packet clumps, while DoHMeter [20] is the feature-extraction module within that toolset, which computes the features and generates the CSV tables. PCAPs allow re-extraction with other tools whilst CSVs are ready-to-use. In this dissertation, we use the official DoHMeter CSVs without any reprocessing of the public PCAPs to train the models integrated into our toolkit. In parallel, our toolkit generates malicious DoH queries and runs a forked DoHMeter to generate CSVs with the identical column schema, theses generated CSVs are unlabelled and used for validation and testing, not for training.

### 2.2.3 Known limitations of the dataset

The public dataset CIRA-CIC-DoHBrw-2020 is highly useful but not free from biases. The benign traffic generated from automated browsing of Alexa Top 10k websites may not reflect enterprise workloads, whilst malicious DoH is generated with only three tunnelling tools (dns2tcp, DNSCat2, Iodine), potentially narrowing the diversity of the dataset [14]. The fixed transmission-rate range for malicious runs (100-1100 b/s) may not cover low-and-slower or bursty regimes of malicious traffic [14]. Consequently, models trained on this corpus may not be efficient against an evasive actor as this dataset has not been generated with evasion in mind. In this dissertation, we train only on the public dataset as released (no reprocessing, rebalancing, or augmentation). Post-training, evasion is assessed with our toolkit: a configurable DoH client that generates malicious traffic while DoHLyzer/DoHMeter passively capture the client ↔ DoH-resolver exchanges and export unlabelled CSVs [15] with the same schema as the public release. These CSVs files are kept strictly out of training and are used solely to stress robustness against evasion.

## 2.3 Evasion methods

Adversaries can shape DoH traffic so that flow-level statistics resemble benign browsing traffic and slip past detectors. There are a few common tactics that may be used and will be described in this section.

### 2.3.1 Low-and-slow rate control

Adversaries deliberately throttle query volumes so that bytes-rates stay near benign browsing. This technique is explicitly recommended for the implementation of a command and control (C2) infrastructure using DNS-Over-HTTPS [21] and is considered when generating datasets such as the CIRA-CIC-DoHBrw-2020 that parameterizes malicious runs with random transmission rates, illustrating the "low and slow" regime discussed in practice [14].

### 2.3.2 Timing jitter

As described by Fox-IT, most C2 (Command and Control) frameworks, that can use DoH as a covert channel expose parameters such as sleep + jitter (e.g. sleep = 10 seconds with 10% of jitter, which gives a random interval around 60 seconds). This weakens strict periodicity rather than eliminating it, so detectors should analyse the distribution of inter-arrival times rather than expect perfectly fixed periods. [22]

### 2.3.3 Chunk sizing

Attackers can split and encode payload into small chunks carried in DNS fields (e.g. QNAME labels or TXT record) so each query is syntactically a valid DNS Query. Because DoH simply tunnels DNS inside HTTPS, the usual DNS size limits still apply (less than or equals to 63 octets per label; less than or equals to 255 octets per FQDN), so the chunk lengths are tuned by the attacker to fit these bounds [3]. Due to the defender's lack of payload visibility with DoH, the data to be analysed shifts from content to side-channels (per-flow bytes/rates, packet-length statistics, inter-arrival timing, and request ↔ response deltas. Chunking contributes to reducing size distributions, diluting volume across many tiny queries, which helps to blur separability against benign browsing. As stated in [23], "the malware reads the exfiltrating data and divides it line by line. Each part of the exfiltrating data is added to the DNS query" which supports the use of chunk sizing for DNS-Over-HTTPS file exfiltration.

### 2.3.4 HTTP method selection

DoH supports both GET and POST methods as stated by RFC 8484 [6]. From an evasion standpoint, POST tends to blend better, as the query is not exposed in the request line/URL in case of an enterprise IDS/Proxy with TLS decryption. In an enterprise context however, GET is mostly privileged for caching reasons, as Google states that POST requests "reduces the cache-ability of responses and can increase DNS latency, so it is not generally recommended", and "Using the GET method can reduce latency, as it is cached more effectively" [24]. However, this evasion method becomes useless if the network does not make use of an IDS/Proxy that uses TLS decryption, as without TLS decryption both methods remain encrypted HTTPS and are largely indistinguishable at content level.

### 2.3.5 DNS-Over-HTTPS Resolver Rotation

Attackers can configure their malicious DoH client with multiple DoH resolvers and rotate across them to evade simple blocklists (IP based blocklists). Real-world tooling shows this explicitly as the ChamelDoH configuration contains "an array of legitimate DoH cloud providers that can be abused for tunnelling," enabling evasion of blocklists (domain-based blocklists or IP blocklists) [18].

From a feature perspective, resolver rotation mainly perturbs metadata and timing.

### 2.3.6 Padding

Padding is a mechanism originally intended to improve the confidentiality of DNS exchanges by making it more difficult to analyse traffic based on packet size. The integration of DoH with http/2 allows the use of the padding mechanisms defined by RFC 7540, which allows the addition of additional bytes in frames to blur the correlation between the length of messages and their content [25]. RFC 8484 specifies that DoH clients can use http/2 padding like any other client in the protocol [6].

In the case of data exfiltration, this padding mechanism can be hijacked by an attacker who chooses to add padding to queries to hide the variability in the size of the requests, thus making them uniform. This technique makes it difficult for detection systems that rely on statistical characteristics, such as average size or the variance of packet lengths. Initially designed as a protective measure against passive analysis, padding thus becomes an evasion tool to more effectively hide malicious activity in DoH traffic.

## 2.4 Detection methods

While DNS-over-HTTPS significantly improves privacy and security for legitimate users, it also complicates the task of defenders, as traditional DNS monitoring and filtering mechanisms lose visibility once queries are encapsulated in HTTPS. Consequently, several detection strategies have been proposed in the literature to identify malicious DoH traffic, ranging from classical signature-based inspection to more advanced anomaly detection and machine learning approaches. These methods shift the focus from payload content, which is encrypted, to side-channel features such as flow-level statistics, timing patterns, and packet size distributions. However, the effectiveness of these approaches is challenged by the evasion methods described in Section 2.3.

### 2.4.1 Signature-based detection

Traditional intrusion detection systems (IDS) rely on signatures (e.g., known bytes patterns, protocol fingerprints) to identify malicious traffic. In the case of DNS-over-HTTPS, this approach is largely ineffective, as the encapsulation within TLS hides both queries and responses from inspection. The only exception is in environments where TLS interception either in enterprise environments or via client-side security software, where a proxy terminates TLS, decrypts traffic, analyse the data and then re-encrypts the connection.

However, studies such as "Killed by Proxy: Analysing Client-end TLS Interception Software" have demonstrated that some antivirus and parental-control applications that perform TLS interception by installing their own root certificates introduces significant security risks [26]. As they were found to be vulnerable to full server impersonation, and some misleadingly present connections as more secure than they actually are. Consequently, even if signature-based detections become technically feasible using TLS Interception, it is a trade-off as it introduces new attack vectors and privacy concerns.

### 2.4.2 Anomaly-based detection

Anomaly detection techniques aim to flag abnormal traffic from a base of benign traffic by examining the statistical properties of flows. The common feature include bytes sent/received, packet-length variance, inter-arrival timing or request-response deltas. For instance, DoHLyzer [15] generates such flow-level representations by extracting statistical features (e.g., packet lengths, inter-arrival times and byte counts). These features are then exported for use in classifiers. The full list of features generated is provided in Appendix A. It has been shown in literature that low-throughput DNS exfiltration channels can be detected by analysing distributional anomalies in flow characteristics [13]. Also, it was demonstrated that DoH tunnels can be identified using a two-layer classification approach based on statistical features and time-series representation of flows [27].

### 2.4.3 Machine Learning models

Also, Machine Learning (ML) models have been also applied to classify DoH traffic, using statistical and time-series features. The use of supervised models such as Random Forest, Decision Trees and Support Vector Machines has shown high accuracy when trained with labelled datasets [14, 27]. For example, a feature-fusion framework was proposed [28], combining multiple feature families, that improved the robustness of detection against covert DoH channel.

While theses ML-based detection systems improve classifications, it remains highly dependent on the quality and representativeness of training data. Models that are trained on public datasets may be overfit to the specific tunnelling tools that were used, and their accuracy can degrade when used against evasive techniques that were describe in Section 2.3.

### 2.4.4 Threshold-based detection models (DoHxP)

In addition to the Machine Learning detection models, threshold-based detection models have been explored as a lightweight way of detecting malicious activity with DNS over HTTPS. These models operate by applying a fixed threshold on statistical feature, such as average packet length, variance of inter-arrival times, or request-response round-trip deltas. The flows that exceed these thresholds are flagged as suspicious, without the need for any complex classifiers.

This approach has been used for DoHxP [17], the authors provided two types of thresholds, each trained on a different dataset. For this dissertation, we will use the "UNB Dataset thresholds" that were generated using the CIRA-CIC-DoHBrw-2020 dataset that we also use for this dissertation, the thresholds and the values used are described in Table 2.

*Table 2. Threshold values defined by DoHxP for detecting malicious DoH traffic (trained on the CIRA-CIC-DoHBrw-2020 dataset)*

| Threshold | Value |
| --- | --- |
| Payload Length | >225 Bytes |
| Traffic volume | >5,000 bytes/s |
| Packet frequency | >150 packets/s |

However, the effectiveness of this detection method is constrained by the variability of benign traffic, some evasion methods such as padding or low-and-slow rate control can deliberately keep the flow's features within the benign ranges. However, this mechanism has the advantage to provide simple and transparent alerts, in contrast to the Machine Learning detection methods.

### 2.4.5 Summary

Table 3 represents a summary of the detection methods described in Section 2, used for detecting malicious DoH traffic.

*Table 3. Summary of detection methods for malicious DoH traffic*

| Approach | Observable features | Typical technique | Representative work |
|---|---|---|---|
| Signature-based | TLS/HTTPS metadata, decrypted payload with TLS interception | Rules/patterns, protocol fingerprints, endpoint blocklists | [26] |
| Anomaly-based | Flow stats (see Appendix A) | Statistics, heuristics | DoHLyzer's features [15], [13], [27] |
| ML-based | Same as anomaly, with time-series windows | Supervised classifiers, requires labelled dataset for training | Dataset [14], [27], [28] |
| Threshold-based (DoHxP) | Simple aggregates per flow, payload length, bytes/s, packets/s | Fixed per-metric threshold, flag if any is exceeded | DoHxP [17] |

# 3. Methods

## 3.1 Introduction

This section presents the methodology that was used to design, implement and evaluate the detection of DNS-over-HTTPS file exfiltration. This covers dataset preparation, feature extraction, machine learning and threshold-based detection models, evasion testing and the setup used. The goal is to provide a reproducible pipeline that train models based on the dataset, captures both benign and malicious DoH traffic, extracts representative flow-based features using DoHLyzer, and runs a predictor that asses the ability of different detection techniques flag the malicious traffic as indeed malicious. This results in a toolkit named DoHExfTlk (DNS-Over-HTTPS Exfiltration Toolkit) [31].

## 3.2 Research Design

We generated and analysed malicious DNS-over-HTTPS traffic using an experimental custom-built pipeline. The pipeline itself focuses on producing realistic exfiltration scenarios under controlled conditions. It allows manipulation of the attack parameters, such as payload encoding, chunk size, compression and timing profiles.

The overall design of the pipeline is illustrated in Figure 2. The architecture shows how malicious DNS-over-HTTPS traffic is generated, captured, and analysed. On the right-hand side, the Docker-based infrastructure hosts the Traefik (used as TLS proxy), the DoH Server (used to process HTTPS DNS queries), and the resolver (used by the DoH server to process the DNS queries). On the left-hand side, the monitoring and analysis components capture the traffic, reconstruct exfiltrated payloads, and extract flow-based features for classification. On top, the DoH Client and Exfiltration Tool makes malicious requests to the DoH server. All these interactions occur within the Internal Docker Network.

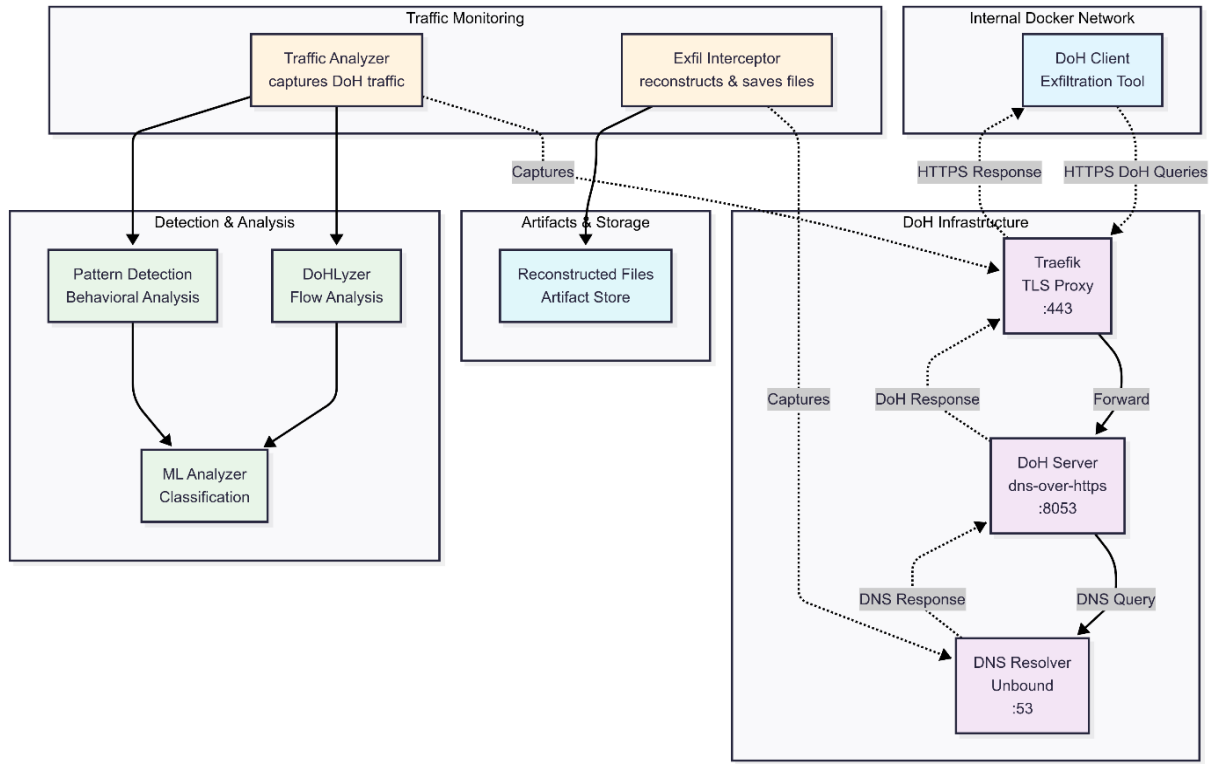

*Figure 2. Overview of the pipeline for DoH exfiltration and detection (https://github.com/AdamLBS/DohExfTlk)*

## 3.3 Dataset preparation

The dataset used in for this project is the CIRA-CIC-DoHBrw-2020 dataset [14] that has already been described in section 2.2. It contains both benign and malicious DoH traffic generated in a controlled environment by the Canadian Institute for Cybersecurity (CIC).

The dataset contains PCAP files as well as pre-processed CSVs files. The CSVs were generated using DoHMeter, a feature extraction tool part of DoHLyzer [15] built on top of the raw PCAP network traces. Each flow in the dataset is thus represented by a set of statistical features described in Appendix A.

For the purposes of this dissertation, the CSV version of the dataset was used for training and evaluating detection models. That dataset was split into training, validation and testing subsets.

## 3.4 Feature extraction and representation

Feature extraction was performed using DoHMeter, a tool provided alongside DoHLyzer [15] that processes raw PCAP network traces into flow-based representations. Each flow is defined as a bidirectional sequence of packets between a client and a resolver within a time window.

In this project, a modified fork [29] of DoHLyzer was used to improve the performance of flow generation and make the tool more resilient to evasion techniques. Two main changes were introduced. First, the garbage collection (GC) stage now occurs in a separated thread, this prevents blocking the main packet processing loop and helps reduce latency during heavy traffic processing, while allowing for automatic cleanup of finished flows each 5 seconds. Secondly the flow clean-up strategy has been optimized in order to handle short-lived flows (e.g., the expiration time for a flow has been reduced from 40 seconds to 20 seconds and the minimum packet count has been reduced from 10 000 to 4 000, if any of these two conditions occurs, the flow is written).

The overall feature extraction process and the implemented modifications are illustrated in Figure 3.

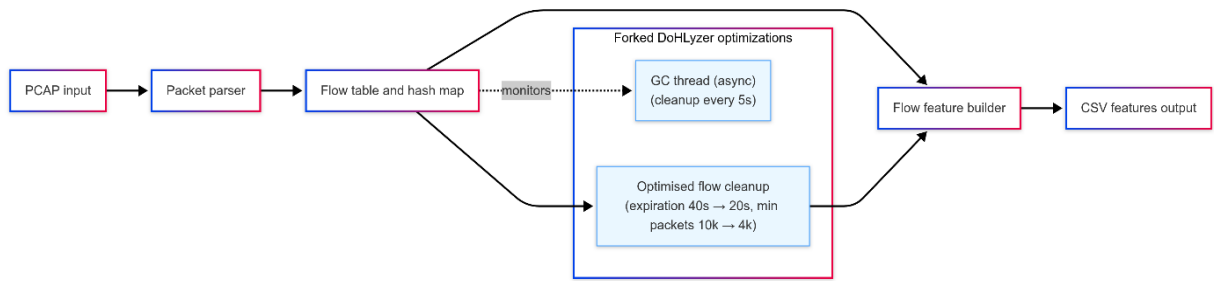

*Figure 3. Feature extraction pipeline in the modified fork of DoHLyzer (https://github.com/AdamLBS/DoHLyzer). The garbage collection was moved to an asynchronous thread executing every 5 seconds, and the flow cleanup strategy was optimized by lowering the expiration threshold (40s -> 20s) and the minimum packet count (10k -> 4k), resulting in more reliable processing of short-lived flows.*

## 3.5 Detection models

Two approaches are used by the toolkit to detect malicious DNS-over-HTTPS traffic: a threshold-based method (DoHxP) [17] and several machine learning classifiers. The toolkit combines both approaches in order to test the effectiveness of each approach in regards to the evasions methods used.

### 3.5.1 Threshold-based detection (DoHxP)

The DoHxP method, as described in section 2.4.4, relies on simple statistical threshold. A flow is flagged as malicious when its features exceed one or more of these thresholds. This approach has the advantage of being lightweight, and doesn't require model training. However, it's also more rigid and is sensitive to the attacker's evasion method. In this project, the threshold rules are defined in a JSON configuration file, located in the models/ directory. This design allows researchers to easily modify or extend the threshold set without changing the codebase. An example configuration file can be seen in Figure 4.

```json
models > {} dohxp_model.json > ...
  1  {
  2    "rules": [
  3      { "feature": "PacketLengthMean", "op": ">", "value": 400, "weight": 0.6 },
  4      { "feature": "PacketTimeVariance", "op": "<", "value": 0.001, "weight": 0.5 },
  5      { "feature": "FlowReceivedRate", "op": ">", "value": 20000, "weight": 0.4 }
  6    ],
  7    "aggregation": "sum",
  8    "clip": [0.0, 1.0],
  9    "bias": 0.0
 10  }
```

*Figure 4. Example JSON configuration file for DoHxP*
*(https://github.com/AdamLBS/DohExfTlk/blob/main/models/dohxp_model.json)*

Each rule specifies a feature, an operator (e.g. > or <), a threshold value, and a weight. The rules are aggregated (in this case by a sum operator), clipped to the [0,1] interval and combined to produce a final score. The aggregated score is compared to a cutoff value (0.5 by default). Flows with scores above or equal to the cutoff are labelled as malicious. The cutoff value can be adjusted to have control over false positives rates.

### 3.5.2 Machine Learning classifiers

The toolkit allows to train several supervised machine learning models on the extracted DoHLyzer features. The training pipeline includes class balancing (SMOTE/undersampling), the probability cutoffs are selected to meet a target FPR (False-positive-Rate) of typically 1%. The implemented models (available in the corpus and the repository inside the "models/" directory) are described in Table 4.

Table 4. Trained machine learning models and artefacts (https://github.com/AdamLBS/DohExfTlk/tree/main/models)

| Model | Artefacts | Description |
|---|---|---|
| Random Forest | models/random_forest.pkl | Tree ensemble classifier |
| Gradient Boosting | models/gradient_boosting.pkl | Boosted trees model |
| Logistic Regression | models/logistic_regression.pkl | Linear baseline model |
| Best Model (CV) | models/best_model.pkl | Model with the best validation score |
| Preprocessing pipeline | models/preprocessors.pkl | Scaling and feature pipeline |
| Metadata | models/metadata.json | Training configuration, hyperparameters, sckit-learn version and validation metrics |
| Thresholds | models/thresholds.json | Probability cutoffs chosen to reach target FPR (false-positive rate) for each model |
| Threshold-based model | models/dohxp_model.json | JSON rules for DoHxP threshold-based model |
| Support Vector Machine | (not provided) | Supported by the code; not distributed as the training time was very expensive. |

In the CIRA-CIC-DoHBrw-2020 dataset, the two files used (l2-benign.csv and l2-malicious.csv) show a clear imbalance in term of size. The malicious file is considerably larger (approximately 155 MB) than the benign file (approximately 11 MB), which indicates that malicious flows significantly outnumber benign flows. Such imbalance can bias the classifier, as they may learn to overfit towards the majority class and misclassify benign flows as malicious, as it was shown in literature [30]. To mitigate this issue, the training pipeline applies class balancing techniques, such as random under sampling of malicious flows and synthetic oversampling of benign flows using the SMOTE (Synthetic Minority Oversampling Technique) algorithm. This technique ensures that both classes are equally represented during training, preventing the classifier from being overfitted and improving its ability to correctly distinguish between benign and malicious traffic.

### 3.5.3 Hybrid approach and reproducibility

A hybrid predictor was implemented to combine both detection paradigms. The predictor tests each available model (whether ML-based or threshold-based) and computes the result for each of them as shown in Figure 5.

```
2025-08-15 08:36:13,415 - INFO - Loaded model: logistic_regression
2025-08-15 08:36:13,534 - INFO - Loaded model: random_forest
2025-08-15 08:36:13,535 - INFO - Loaded DoHXP rule-based model from ../models/dohxp_model.json
2025-08-15 08:36:13,535 - INFO - Loading data: /app/captured/output.only_172.18.0.5.csv
2025-08-15 08:36:13,548 - INFO - Features used: 31/31
2025-08-15 08:36:13,548 - INFO - === PREDICTIONS ===
2025-08-15 08:36:13,556 - INFO - 🤖 LOGISTIC_REGRESSION:
2025-08-15 08:36:13,556 - INFO -     - Benign: 1
2025-08-15 08:36:13,556 - INFO -     - Malicious: 0
2025-08-15 08:36:13,556 - INFO -     - Threshold applied: 0.957
2025-08-15 08:36:13,556 - INFO -     - Avg confidence: 0.996
2025-08-15 08:36:13,586 - INFO - 🤖 RANDOM_FOREST:
2025-08-15 08:36:13,586 - INFO -     - Benign: 1
2025-08-15 08:36:13,586 - INFO -     - Malicious: 0
2025-08-15 08:36:13,586 - INFO -     - Threshold applied: 0.209
2025-08-15 08:36:13,586 - INFO -     - Avg confidence: 0.980
2025-08-15 08:36:13,586 - INFO - 🤖 DOHXP:
2025-08-15 08:36:13,587 - INFO -     - Benign: 0
2025-08-15 08:36:13,587 - INFO -     - Malicious: 1
2025-08-15 08:36:13,587 - INFO -     - Threshold applied: 0.500
2025-08-15 08:36:13,587 - INFO -     - Avg confidence: 1.000
2025-08-15 08:36:13,587 - INFO -
=== SUMMARY OF PREDICTIONS ===

Model                 Benign    Malicious    Total    Threshold    Confidence
--------------------------------------------------------------------------------
logistic_regression      1          0          1        0.957        0.996
random_forest            1          0          1        0.209        0.980
dohxp                    0          1          1        0.500        1.000
```

*Figure 5. Example predictor output showing per-model results (benign vs malicious flows, threshold applied, and average confidence), and a consolidated summary of predictions*

To ensure reproducibility of the experiments, all trained models and preprocessing steps were exported as pkl artefacts. The .pkl files correspond to serialized Python objects produced by the training pipeline. They are created by the joblib library, which allows complex objects such as models to be saved to disk and reloaded later without retraining. In this project, each "/models/*.pkl" file stores a trained classifier (e.g., Random Forest, Gradient Boosting, Logistic Regression) along with its internal parameters.

## 3.6 Evasion testing methodology

### 3.6.1 Exfiltration Client

To evaluate the robustness of detection models, different evasion scenarios were designed and executed using the exfiltration client implemented in the toolkit (exfiltration/client/run_client.py). This client accepts configuration files in the JSON format.

The exfiltration client follows a structured workflow to transmit files over DNS-over-HTTPS, as shown in Figure 6. The workflow begins with a connectivity check, where the client verifies that the configured DoH resolver is reachable by sending DNS queries for well-known domains to the DoH resolver that will be used for exfiltration. Once validated, the file to be exfiltrated is processed through a data preparation pipeline, that may include (depending on the configuration chosen by the user) compression, encryption and encoding (e.g., base64 or base32).

The resulting payload is then split into chunks that are embedded into DNS query subdomains along with a session identifier and sequence number. Each chunk is transmitted as a DoH query over HTTPS, following the timing pattern defined in the configuration and failed queries are retried until successful.

Finally, the client outputs a log file that includes the transmission success rate, throughput, retries and duration of the communication.

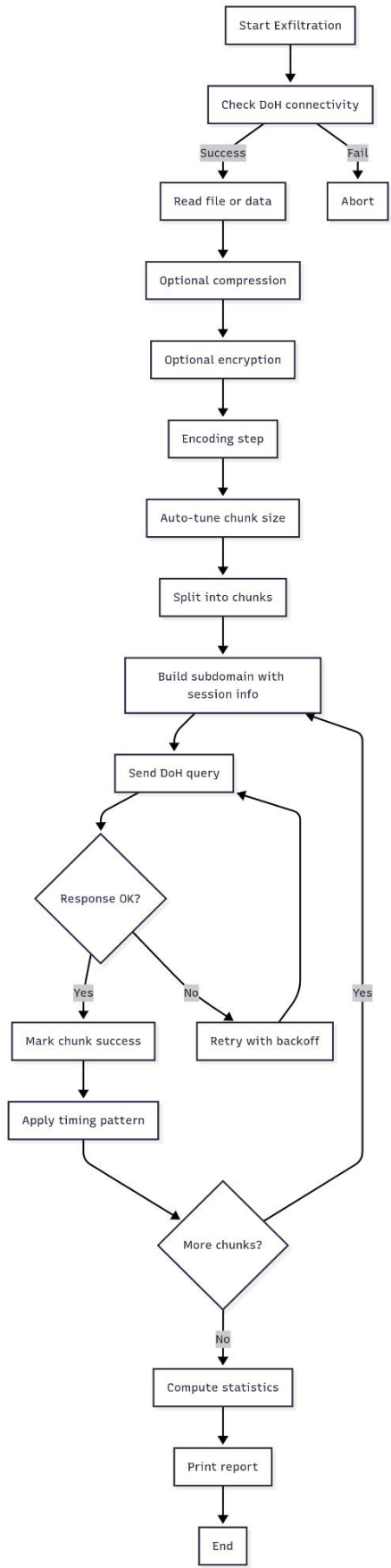

*Figure 6. Exfiltration Client Workflow*

## 3.6.2 Exfiltration Server

The exfiltration server as highlighted on the right side of Figure 7 is essential to ensure the exfiltration occurs and to simulate real-world situations. Within the toolkit, this component plays a dual role that can be leveraged by both red teams and blue teams. From the red team perspective, the server validates that payload sent through the DNS-over-HTTPS covert channel are successfully reconstructed on the attacker side, demonstrating the feasibility of data exfiltration under different evasion profiles.

From a blue team perspective, the server provides labelled evidence of malicious DNS activity, including the reconstructed payload, statistics and timing information. These can be used to train a detection model, or to enrich incident response literature and better understand how adversarial behaviours manifest in network traces.

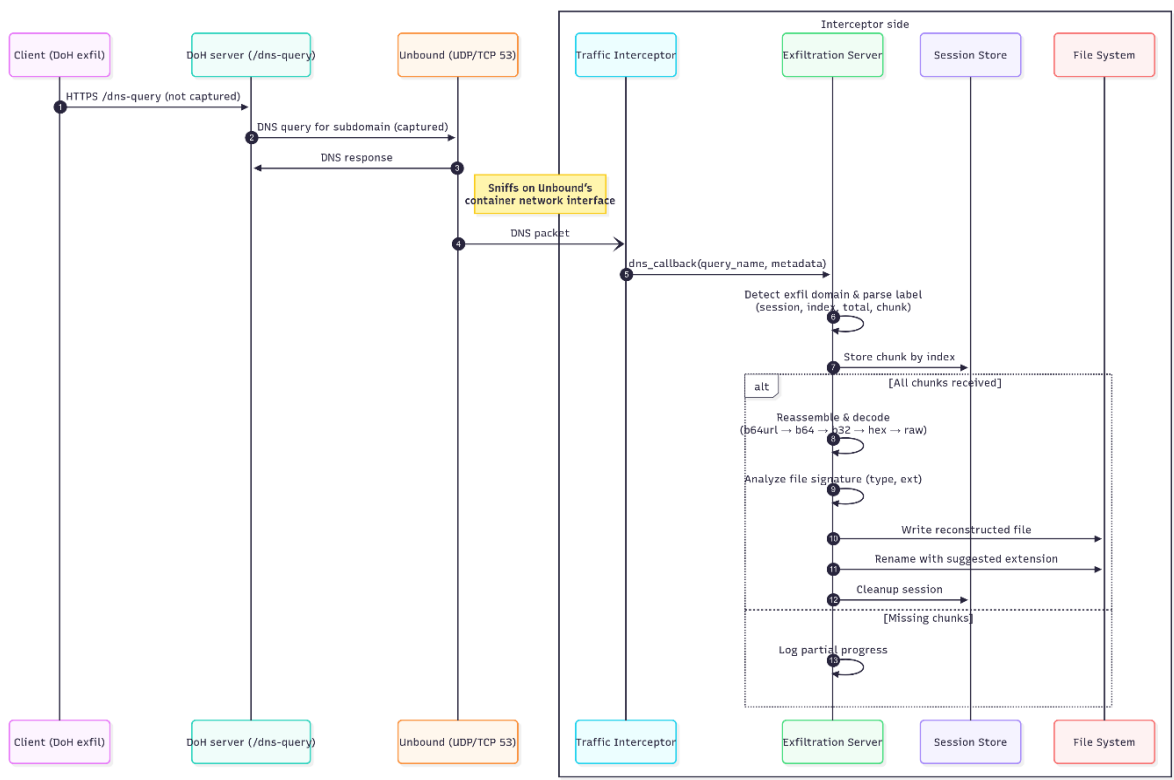

*Figure 7. DNS-Over-HTTPS file exfiltration server*

## 3.6.3 Configuration files

Each exfiltration scenario is defined through a JSON configuration file, where parameters such as chunk size, encoding, timing, and domain rotation can be specified. This modular design makes it easy to reproduce different attacker behaviours. An example configuration is shown in Figure 8.

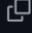

*Figure 8. Example JSON configuration file for a low-speed exfiltration profile (https://github.com/AdamLBS/DohExfTlk/blob/main/exfiltration/client/test_configs/low_speed.json)*

### 3.6.4 Configuration generator

In order to allow researchers and users to easily generate configuration profiles and reduce the risk of misconfiguration, a configuration generator is present in the toolkit, allowing to automatically generate configuration profiles. An example of the generator in use is shown in Figure 9.

```
(.venv) → client git:(main) ✗ python config_generator.py --create
Creating new DoH evasion configuration
DoH Evasion Configuration Generator
=====================================================
Configuration name: Test
Description: Test

Exfiltration configuration:
DoH Server [https://doh.local/dns-query]:
Target domain [exfill.local]:
Chunk size [30]: 14
Available encoding: base64, hex, base32, custom
    Selection [base64]: base64
Timing patterns: regular, random, burst, stealth
    Selection [regular]: random
Base delay in seconds [0.2]: 0.4

Evasion options:
Compression [y/N]: y
Encryption [y/N]: y
Encryption key: test
Subdomain randomization [Y/n]: y
Padding [y/N]: y
Domain rotation [y/N]: n
Delay variance [0.1]: 0.4

Configuration saved: test_configs/test.json
```

*Figure 9. Example execution of the interactive configuration generator.*

The user specifies parameters such as encoding, chunk size, timing pattern and evasion options, and the tool saves the corresponding JSON file.

### 3.6.5 Evasion profiles

To assess the robustness of detection methods, several exfiltration profiles were designed using the JSON configuration system. Each profile corresponds to a different attacker strategy, inspired by those described in Section 2.3. The following profiles described in Table 4 were used in this dissertation (and comes shipped with the toolkit).

*Table 5. Overview of evasion profiles used in the experiments*

| Profile | Key parameters / techniques | Attacker strategy |
|---|---|---|
| Classic exfiltration | Base 64 encoding, medium chunk, regular timing | Basic attack, without any evasion |
| low-and-slow | Very small chunks, long delays, random timing | Stealthy strategy, privileging stealthiness over transfer speed |
| Burst | Large chunks, high throughput in short bursts, idle intervals | Overwhelm detectors with peaks of activity |
| Stealth / Randomised | Irregular chunk sizes, padding, compression, subdomain randomisation, high delay variance | Mimic legitimate traffic |
| Speed | Large chunk, no intervals, continuous transmission | Maximise throughput, prioritising speed over stealth |

By executing these different profiles multiple times with varying payload (text, image, or even binary files) of different size, the models are tested against a wide spectrum of attacker behaviours. This ensures that both threshold-based and machine learning detection models are evaluated not only on standard exfiltration attempts, but also on adversarial strategies that are more difficult to detect. In this way, the toolkit provides a robust assessment of the resilience and limitations of the proposed detection approaches.

## 3.7 Experimental setup

### 3.7.1 Host Infrastructure

All experiments were executed on a virtual machine provided by the University of Kent. The VM was running Ubuntu 24.04.2 LTS with the specifications detailed in Table 6.

*Table 6. Traffic generation and feature extraction infrastructure specifications*

| Component | Specification |
| --- | --- |
| OS | Ubuntu 24.04.2 LTS |
| RAM | 8 GB |
| Storage | 80 GB SSD NVME |
| CPU | Intel Haswell 4vCore @ 2.6 GHz |

The model training on the other side was carried out on the University of Kent high-performance computing cluster, to speed up the training. The training jobs were run on the CPU partition of the cluster, with access to 40 CPU cores and 80GB of RAM, using a slurm-compatible bash script as shown in Figure 10. Although the cluster also provides GPU nodes, only CPU resources were used in this project, as GPU acceleration has not been yet implemented in the toolkit.

```
  GNU nano 6.2
#!/bin/bash
#SBATCH --job-name=doh_rf_full
#SBATCH --partition=cpu
#SBATCH --ntasks=1
#SBATCH --cpus-per-task=40
#SBATCH --mem=80G
#SBATCH --output=log_rf_full_%j.out

source /home/hmc/ae469/Kent-Dissertation/.venv/bin/activate
cd /home/hmc/ae469/Kent-Dissertation/ml_analyzer
python model_trainer.py
```

*Figure 10. Slurm script used to run the model_trainer Python script*

### 3.7.2 Docker Infrastructure

The experimental environment was deployed as a set of Docker containers connected through an internal Docker network (dohnet). Each major component of the pipeline (clients, proxy, DoH server, resolver, exfiltration server, and traffic analyzer) was containerised to ensure portability and reproducibility of the experiments. A "client_test" container was also added to facilitate access to the Docker internal network for further research or experiments. The infrastructure is described in Figure 11.

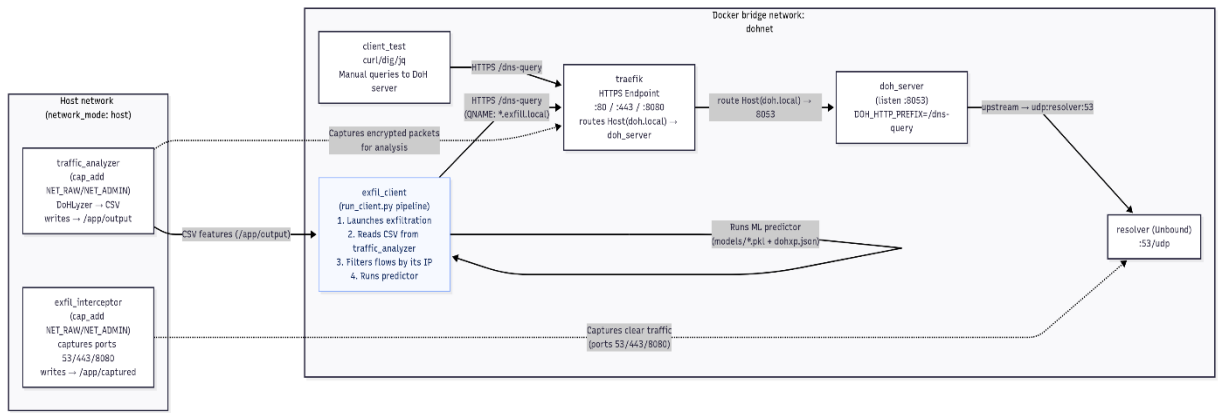

*Figure 11. DoHExfTlk Docker Infrastructure*

Each container in this toolkit has a specific role within the pipeline. Their purpose, network configuration and exposed ports are summarized in Table 7.

*Table 7. Containers deployed in the Docker toolkit (https://github.com/AdamLBS/DohExfTlk/blob/main/docker-compose.yml)*

| Container Name | Role | Network / Ports |
| --- | --- | --- |
| traefik | Reverse proxy and TLS termination. Exposes the DoH (DNS-over-HTTP) server endpoint (doh.local) over HTTPS | Host: 80, 443, 8080 (Traefik Dashboard) |
| doh_server | DNS-over-HTTP docker image [32] using m13253's dns-over-https server [33]. Listens on port 8053 and forwards to resolver | Local Docker Network (dohnet) |
| resolver | Upstream DNS resolver using Unbound [34], answers DoH server queries on UDP/53 | Local Docker Network (dohnet) |
| client_test | Utility client running Ubuntu, for manual DoH query, used as an endpoint for further research | Local Docker Network (dohnet) |
| exfil_client | Runs the research pipeline: generates DoH exfiltration traffic, collects analyzer output, filters flows, and executes the predictor | Local Docker Network (dohnet) |

| exfil_interceptor | Packet sniffer that captures cleartext traffic from the resolver, reconstructs the exfiltrated data and writes to /app/captured | Host network (raw capture) |
|---|---|---|
| traffic_analyzer | Packet sniffer that runs the DoHLyzer's fork [29], sniffs the traffic from Traefik, generates flows and exports CSV to /app/output | Host network (raw capture) |

### 3.7.3 Directory and volume mapping

To ensure reproducibility and facilitate data exchange between the containers, several Docker volumes were mounted from the host into the containers. For example, these volumes allow the exfiltration client to consume outputs from the analyzer, while having access to the trained models and execute the predictor as shown in Figure 12.

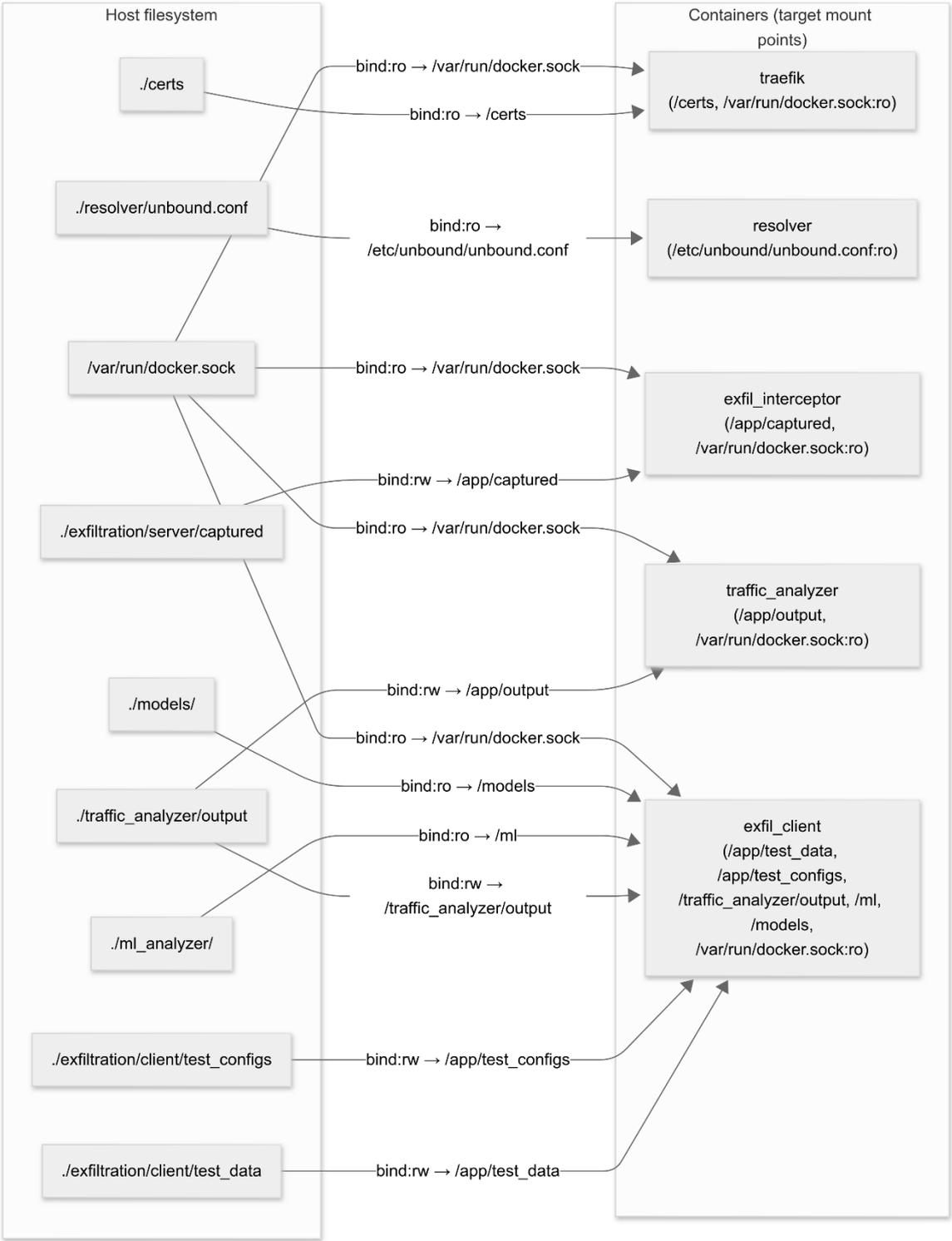

*Figure 12. Directory and volume mapping between the host and the containers for the DoHExfTlk toolkit.*

As shown in Figure 12, the host's Docker socket is deliberately exposed to some containers, whilst this is a known security risk [35] (and part of the reason for which this toolkit is not made for use in production environments), it enables some containers such as exfil_interceptor and traffic_analyzer to dynamically gather the network interfaces of target containers they want to sniff (e.g. Traefik, resolver).

In a hardened setup, this capability should be replaced by a restricted mechanism such as a read-only API with scoped credentials, or a predefined interface bindings to avoid granting broad access to the Docker daemon.

## 3.8 Evaluation metrics

The evaluation of the detection models in this project are performed at two levels. First the success of the exfiltration process itself, and second the accuracy and reliability of the detection pipeline. While the exfiltration metrics provide the ground truth of whether data exfiltration was successfully achieved, the detection metrics represent the predictions of the detection models. This dual view allows for a direct mapping between actual exfiltration attempts and their detection outcomes.

### 3.8.1 Exfiltration-level metrics

The exfiltration client records statistics for each test, including the total duration of the transfer, the number of chunks sent, number of retransmissions, success rate of the transmission and throughput calculated in bytes per second.

These values provide insights into the efficiency of different exfiltration configurations (e.g. low-and-slow against burst). For instance, the client logs the percentage of chunks transmitted, the achieved throughput and whether any retries have been made, as shown in Figure 13.

```
exfiltration > client > results > run-20250813-131727 > burst > logs > ≡ client.log
702  [656/671] Sent chunk: 17550910-0655-0671-teoXuvIq6koYyK0wB5X8sMYDP27uNf...
703  [657/671] Sent chunk: 17550910-0656-0671-CbrpXSVIG_BiI7JSsKPnVOW0B3XLgU...
704  [658/671] Sent chunk: 17550910-0657-0671-eRJPvK35zJqVTHGobo1M-8lt9X66tz...
705  [659/671] Sent chunk: 17550910-0658-0671-6aOwFK9hqjJP5Jee6hfdqBS8DLw5Rh...
706  [660/671] Sent chunk: 17550910-0659-0671-_4N68iYL4JPu6U-dP-Wm-P_YO3oQGs...
707  [661/671] Sent chunk: 17550910-0660-0671-lYxtOyZwK1j9AwZ3p5UqOFzHCJLgcd...
708  [662/671] Sent chunk: 17550910-0661-0671-2fxN-Xz9lQ-o0u1JvpIcWYS4zUrJNX...
709  [663/671] Sent chunk: 17550910-0662-0671-h70cSaCsJEmt1EAwlebL6rtKbxqcfp...
710  [664/671] Sent chunk: 17550910-0663-0671-FPR1tJLLF7_SXOnlbMTcX11L3mFXcw...
711  [665/671] Sent chunk: 17550910-0664-0671-Zo3uSXu9wvnf0p-d91U9tsiMLLAvcU...
712  [666/671] Sent chunk: 17550910-0665-0671-us3b4_wKJ5t4PmhOXrU9nXdfgtMZnw...
713  [667/671] Sent chunk: 17550910-0666-0671-qHjeb9jG1JgDk0j5hEj_nTsSr_UkPg...
714  [668/671] Sent chunk: 17550910-0667-0671-Y1tok1VlU9fuupLvFe8GjNPYD23ycR...
715  [669/671] Sent chunk: 17550910-0668-0671-P8PFkQCAn6un-C_cuYSXoP-fflQgo4...
716  [670/671] Sent chunk: 17550910-0669-0671-VM1GFNiGXCfzoaBsgyNfsX_wfehigB...
717  Progress: 99.9% (670/671) - ETA: 0.1s
718  [671/671] Sent chunk: 17550910-0670-0671-fj0AAA==...
719
720  Transmission complete: 100.0% success rate
721  Actual time: 44.3s (estimated: 39.5s)
722
723  📈 EXFILTRATION STATISTICS:
724    Duration: 44.33 seconds
725    Total chunks: 671
726    Successful: 671
727    Failed: 0
728    Retries: 0
729    Success rate: 100.0%
730    Throughput: 355.11 bytes/sec
731    Total bytes: 15742
```

*Figure 13. Example of logs for the client (https://github.com/AdamLBS/DohExfTlk/blob/main/exfiltration/client/results/run-20250813-131727/burst/logs/client.log)*

### 3.8.2 Detection-level metrics

Once the traffic is captured and converted into flow-based features by the forked DoHLyzer meter [29], the predictor [38] applies the available models in the /models/ directory (such as the machine learning classifiers and DoHxP) and outputs the value for the metrics as shown in Table 8.

*Table 8. Metrics reported by the predictor for each model*

| Metric | Description |
| --- | --- |
| Benign / Malicious | Number of flows classified into each category (simplified confusion matrix) |
| Threshold applied | Probability cutoff used for the classification |
| Average confidence | Mean probability score assigned by the model to its prediction |

These metrics are essential to assess the model's behaviour against different exfiltration scenarios, as they allow both a comparison across models and a deeper understanding of their prediction confidence.

*Figure 14. Example of the predictor's output (https://github.com/AdamLBS/DohExfTlk/blob/main/exfiltration/client/results/run-20250815-160144/big-burst/logs/predictor_big-burst-1755273716.log)*

## 3.9 Summary

This section presented the methodology adopted throughout the project. The exfiltration toolkit was first presented, showing how DoH-based data exfiltration scenarios were generated, whilst traffic capture and transformation into flow-based features was performed using the forked DoHLyzer meter, ensuring reproducibility and consistency across experiments.

The test-lab relies on a fully containerised architecture using Docker Compose. Dedicated containers were provisioned for the DoH server, resolver, client exfiltration tool

The predictor was then introduced as the component applying both machine learning models and threshold-based models to the malicious flows generated by the toolkit.

## 3.10 Continuous integration (CI/CD)

To ensure repeatability and up-to-date documentation, the toolkit includes three GitHub Actions workflows in its repository [31], that run automatically on pushes and pull requests to the main branch.

### 3.10.1 Docker image build

This workflow ensures that any contributor can detect any early failure if any Dockerfile or dependency regresses, which guarantees anyone can reproduce the build. As shown in Figure 15, the workflow checkouts the repository, prepare directories expected by containers, generates certificates and builds images for all the containers. It is triggered when any changes occur in the main branch of the repository.

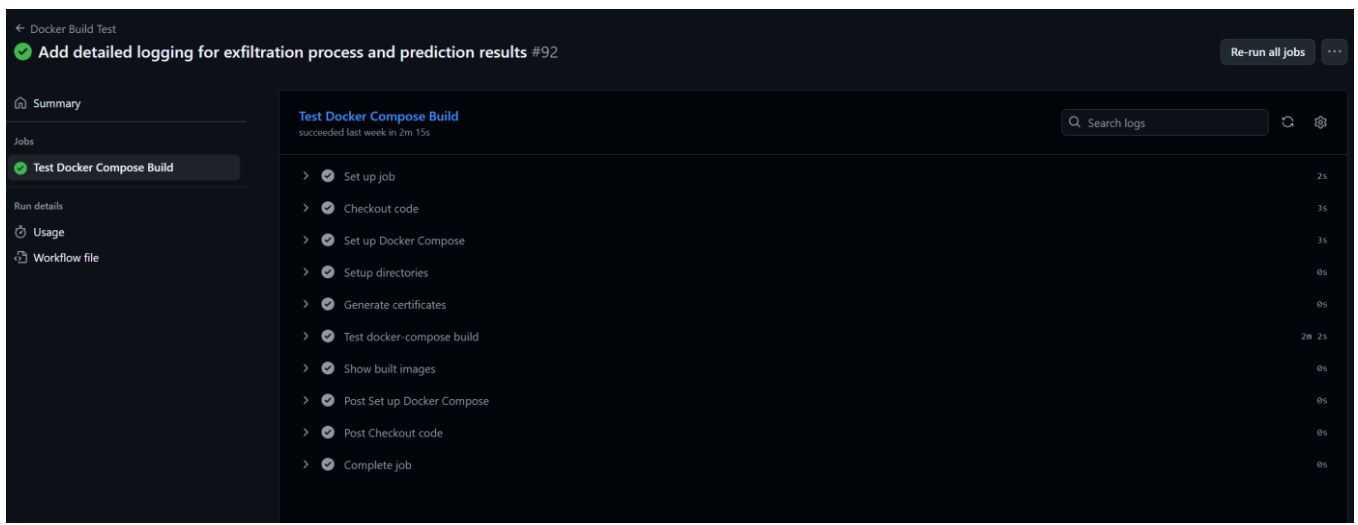

*Figure 15. Example of the Docker Build Test workflow being ran (https://github.com/AdamLBS/DohExfTlk/blob/main/.github/workflows/action.yml).*

### 3.10.2 Publish Docs to GitHub Pages

This workflow ensures that each update of the documentation is deployed and pushed on the website that hosts the documentation [36], which guarantees the freshness of the documentation. It is triggered for every push that edits any of the README.md, docs/**, mkdocs.yml files present in the repository. As shown in Figure 16, the workflow builds the documentation using mkdocs [37] then deploys the generated documentation on GitHub Pages, that is accessible on the dohexftlk.admlbs.fr domain.

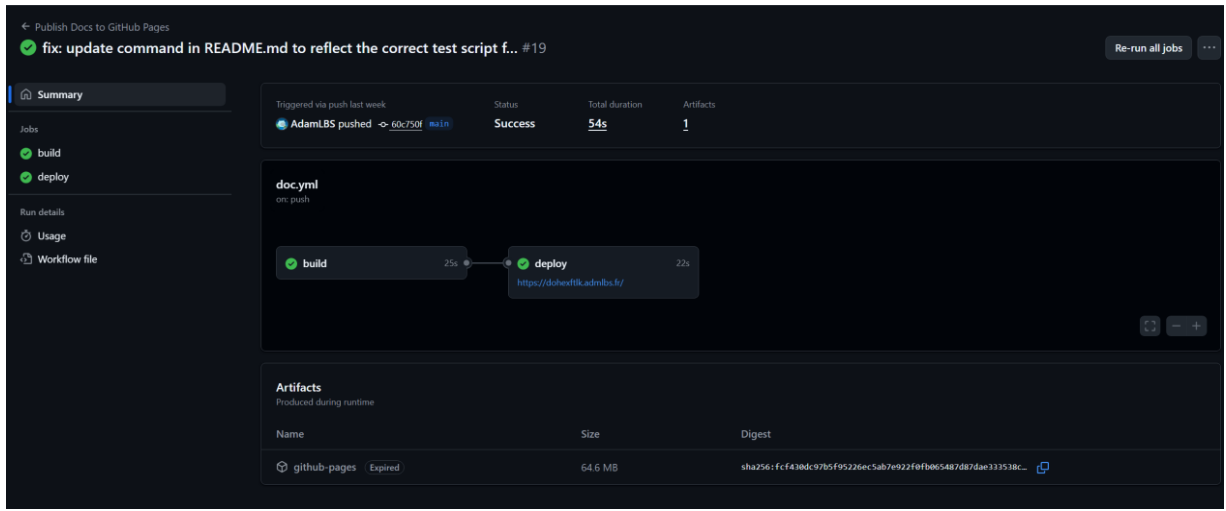

*Figure 16. Example of the Publish Docs to GitHub Pages workflow being ran (https://github.com/AdamLBS/DohExfTlk/blob/main/.github/workflows/doc.yml).*

### 3.10.3 Mirror to Kent GitLab

This workflow ensures that the mirrored repository available in the University of Kent GitLab is up to date. It is run on each push made on main, which makes the toolkit available internally at the University-level. Figure 17 shows the workflow being ran, by retrieving a SSH private key stored as a secret, GitHub can push the changes to the mirrored repository.

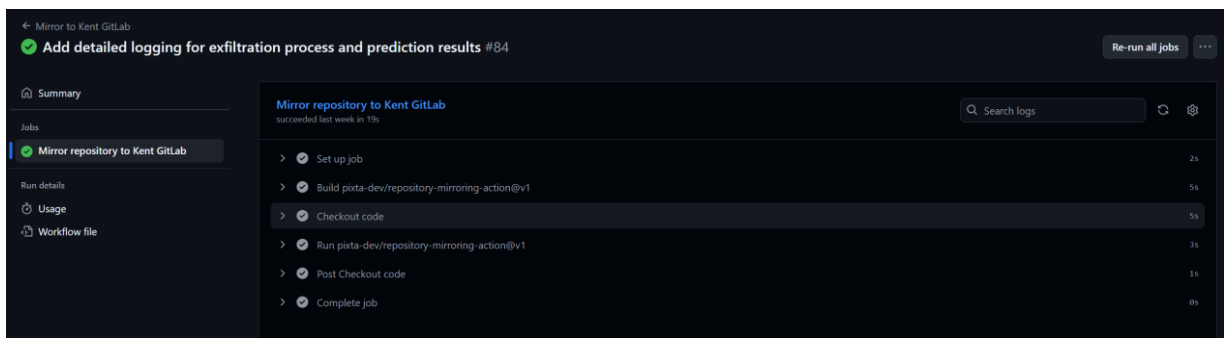

*Figure 17. Example of the Mirror to Kent GitLab workflow being ran (https://github.com/AdamLBS/DohExfTlk/blob/main/.github/workflows/mirror.yml)*

# 4. Results

This section presents the results obtained from the experiments conducted with the DoHExfTlk pipeline and the trained predictor. This evaluation covers two aspects. First scenario-based exfiltration runs that emulate different attacker strategies (big-burst, burst, classic, low-speed, speed, stealth). Secondly, the performance of the detection models when applied to labelled benign and malicious traffic from the dataset.

## 4.1 Baseline evaluation on labelled dataset

Before testing the detection models against custom exfiltration scenarios, we first validated them on the CIRA-CIC-DoHBrw-2020 labelled dataset used to train the models in the first place [14].

Table 9 presents the performance of the four detections approaches provided with the toolkit and tested against a CSV file (l2-malicious.csv) containing only malicious requests: Gradient Boosting, Logistic Regression, Random Forest and the threshold-based DoHxP method. The values correspond to the number of flows predicted as benign or malicious, the detection threshold applied, and the average prediction confidence.

| Model | Benign | Malicious | Total | Threshold | Confidence | Overall malicious detected requests |
|---|---|---|---|---|---|---|
| Gradient Boosting | 7 | 249829 | 249836 | 0.001 | 1.000 | 99.997% |
| Logistic Regression | 157389 | 92447 | 249836 | 0.961 | 0.907 | 37.0% |
| Random Forest | 14 | 249822 | 249836 | 0.024 | 1.000 | 99.994% |
| DoHxP | 244637 | 5199 | 249836 | 0.500 | 0.983 | 2.1% |

*Table 9. Baseline prediction results of the four detection models tested on the malicious-labelled dataset l2-malicious.csv (Corpus: Dataset/CSVs/Total_CSVs.zip)*

We also evaluated the predictor against a CSV file (l2-benign.csv) containing only benign requests, to evaluate how well the detection models avoid false positives. Table 10 presents the results. These values correspond to the number of flows predicted as benign or malicious, the applied threshold, the average prediction confidence, and the overall rate of false positive. To compute the FPR (False Positive Rate) we use this simple mathematical formula:

$$FPR = \frac{False\ Positives}{Total\ number\ of\ benign\ requests} \times 100$$

| Model | Benign | Malicious | Total | Threshold | Confidence | Overall false positives |
|---|---|---|---|---|---|---|
| Gradient Boosting | 19592 | 215 | 19807 | 0.001 | 1.000 | 1.1% |
| Logistic Regression | 19620 | 187 | 19807 | 0.961 | 0.936 | 0.9% |
| Random Forest | 19674 | 133 | 19807 | 0.024 | 0.024 | 0.7% |
| DoHxP | 14416 | 5391 | 19807 | 0.500 | 0.500 | 27.2% |

*Table 10. Baseline prediction results of the four detection models tested on the benign-labelled dataset l2-benign.csv (Corpus: Dataset/CSVs/Total_CSVs.zip)*

The baseline evaluation highlights significant differences between machine learning models and threshold-based detection. Gradient Boosting and Random Forest achieved near-perfect detection of malicious traffic ($\cong 99.99\%$) but at the cost of higher false positive rates ($\cong 1\%$). Logistic Regression on the other hand performed in a more balanced way, correctly classifying a significant portion of benign traffic (99.1% accuracy on benign dataset) but with reduced sensitivity on malicious traffic (37% of accuracy). DoHxP, the threshold-based detection method proved itself to be ineffective in both cases. It detected only 2.1% of malicious requests while also misclassifying more than a quarter of benign traffic (27.2%).

Overall, these results confirm that while machine learning models provide stronger detection capabilities than thresholds, each model comes with trade-offs. Gradient Boosting and Random Forest show strong potential for high recall but may need calibration (or a more diverse dataset to be trained on) to reduce false positives, whereas Logistic Regression is less sensitive but more precise on benign traffic.

## 4.2 Overview of scenarios

After validating the models on labelled data, we now evaluate their performance under controlled exfiltration scenarios with the DoHExfTlk toolkit.

Six configurations were designed to emulate different adversarial strategies, based on those described in Section 2.3. Ranging from simplistic high-volume attacks to stealth techniques. These scenarios vary in terms of chunk size, encoding, compression, timing patterns.

Table 11 summarizes the six scenarios that were tested: Big-burst, burst, classic, low-speed, speed and stealth. Together they cover a broad spectrum of attacker behaviours, allowing us to assess how each detection model reacts to different evasion strategies.

| Scenario | Chunk size / Encoding | Timing pattern | Key features | Purpose |
| --- | --- | --- | --- | --- |
| Big-burst | 30 / base64 | Burst (100 queries, 0.1s) | No compression, no evasion | High-volume exfiltration |
| Burst | 30 / base64 | Burst (10 queries, 0.5s) | Compression enabled | Shorter burst with compression |
| Classic | 40 / base64 | Regular (0.5s) | No compression, no evasion | Naïve configuration |
| Low-Speed | 15 / base32 | Random (average 2.5s, with a variation of 5s) | Compression, domain rotation, subdomain randomisation, padding | Low-and-slow stealthy exfiltration |
| Speed | 60 / base64 | Burst (20 queries, 0.001s) | Compression | Tests maximum throughput |
| Stealth | 12 / base64 (custom, with URL-safe) | Stealth ($8s \pm 3s$) | Compression, encryption, padding, subdomain randomization, domain rotation | Targets ML detection |

*Table 11. Overview of the six exfiltration scenarios configured with the DoHExfTlk client, highlighting key parameters and their intended purpose (https://github.com/AdamLBS/DohExfTlk/tree/main/exfiltration/client/results/run-20250815-160144)*

## 4.3 Results by detection model

This subsection compares the performance of the four detection approaches across the different exfiltration scenarios (big-burst, burst, classic, low-speed, speed, stealth) using an image file as the file to be exfiltrated (seen in Appendix 2). The analysis highlights their ability to flag malicious traffic under different evasion configurations, as well as their limitations in terms of false negatives or misclassifications.

Table 12. Detection rates (%) and number of flows detected for each model across the six exfiltration scenarios (https://github.com/AdamLBS/DohExfTlk/tree/main/exfiltration/client/results/run-20250815-160144)

| Scenario | Gradient Boosting | Random Forest | Logistic Regression | DoHxP |
|---|---|---|---|---|
| Big-burst | 100% (1/1) | 100% (1/1) | 0% (0/1) | 100% (1/1) |
| Burst | 100% (1/1) | 100% (1/1) | 0% (0/1) | 100% (1/1) |
| Classic | 100% (6/6) | 100% (6/6) | 0% (0/6) | 100% (6/6) |
| Low-speed | 100% (37/37) | 100% (37/37) | 0% (0/37) | 100% (37/37) |
| Speed | 100% (1/1) | 100% (1/1) | 0% (0/1) | 100% (1/1) |
| Stealth | 99% (253/255) | 99% (253/255) | 7% (18/255) | 100% (255/255) |

As shown in Table 12, both Gradient Boosting and Random Forest achieved near-perfect detection across all scenarios, consistently flagging malicious traffic with very few misses. Which is in line with the baseline results of Section 4.1.

## 4.4 Interpretation

The results obtained across the different exfiltration scenarios show that machine learning models, such as Gradient Boosting and Random Forest, provide the most reliable detection capabilities, even against stealthy traffic. By contrast, Logistic Regression was unable to properly detect the malicious flows, while DoHxP achieved perfect recall but at the cost of rigidity.

It is worth noting that Logistic Regression, as linear model, assumes that malicious and benign traffic can be separated by a simple linear boundary. As such, this assumption rarely holds in the context of encrypted network traffic, where patterns are complex and nonlinear. As stated in literature, linear and logistic regression models are constrained by their reliance on separability [39, 40]. This explains why Logistic Regression underperformed in our experiments, while other methods such as Random Forest were able to capture these non-linear relationships and achieve a more reliable detection.

However, these promising results must be interpreted with caution. The experiments were conducted in a controlled environment where only benign traffic from the reference dataset and malicious traffic generated by a prototype of the toolkit were present. In such settings, separation between classes is clearer and the results may appear more robust than they would be in a noisy environment (e.g., real-world traffic). In practice, benign variations of DoH usage (e.g., browsers requests or high-volume DNS activity) could be misclassified as exfiltration, especially by threshold-based methods such as DoHxP (which has been shown to have the highest FPR rate against the benign labelled dataset).

Also, the models were trained and validated on the public CIRA-CIC-DoHBrw-2020 dataset [14]. While this ensures reproducibility, it also risks overfitting to specific feature distributions. As a result, the actual detection rate in the wild is likely to be lower.

Another factor that influenced the results is the flow generation pattern of the scenarios themselves. Very fast configurations such as burst and big-burst produced only a single flow due to their speed. This makes them straightforward to classify, as the entire exfiltration is concentrated in one strongly anomalous instance. In contrast, stealthier configurations such as stealth spread the exfiltration across hundreds of smaller flows, which both increases the detection difficulty and better mimics realistic attacker behaviour. This explains why the models achieved perfect accuracy in the fast scenarios but exhibited a small number of misclassifications in stealth configuration.

Overall, the interpretation of these results is that machine learning models are highly effective for detecting DNS-over-HTTPS exfiltration in controlled conditions, but their performance should not be overestimated in real-world conditions. While threshold-based methods such as DoHxP can complement machine learning detection methods, they risk generating false positives when applied to diverse benign traffic, as shown in Section 4.1.

# 5. Discussion

## 5.1. Summary of findings

As stated in Section 4, two patterns emerge. First, the tree-based models (Gradient Boosting and Random Forest) show robustness against adversarial scenarios and were accurate when tested against benign/malicious baselines. Whereas linear models such as Logistic Regression, struggle to correctly classify the benign/malicious baselines flows and proved itself ineffective against adversarial scenarios. Also, threshold-based models such as DoHxP proved itself accurate against each scenario tested with the toolkit. However, this must be put into perspective, as it showed a lot of false positives, having a higher false positive rate than false negative rate. Mainly because using fixed thresholds is overly restrictive and generalise poorly to broader traffic.

## 5.2 Toolkit contribution

This dissertation's practical contribution is the DoHExfTlk toolkit [31], an all-in-one containerised testbed that automatize the study of DNS-over-HTTPS exfiltration. It combines configurable traffic generation, controlled interception/capture, feature extraction, multi-model prediction, into a single and reproducible pipeline.

Table 13 shows the main features provided by the toolkit.

*Table 13. Main features of the DoHxP toolkit*

| Component | Capability | Key options | Value |
|---|---|---|---|
| Exfiltration Client | Evasion-aware file-exfiltration traffic generation | Encoding (base64, base32/base64-urlsafe), chunk size, compression, encryption, padding, subdomain randomisation, resolver rotation, timing | Emulates realistic attacker behaviour, |
| Traffic Interceptor (exfiltration server) | Resolver-side capture | Automatically find the right network interface to listen to Reconstruct files exfiltrated via DoH | Emulates realistic attacker behaviour, on the attacker side, mimics an exfiltration server |

| Feature extraction (with DoHLyzer's fork) | Flow building & feature extraction | Short-flow handling | Stable features for model comparisons |
|---|---|---|---|
| Predictor | Unified multi-detector engine | Handles multiple ML models based on .pkl files, handle threshold-based models with JSON configuration | Applies predictions based on the DoHLyer flow, for model comparison |
| Docker | Reproducibility, portable, automated runs | Docker Compose with a containerized setup allows for automated runs, CI build/tests, bash scripts for pipeline runs and allow users to select the container they want to run (e.g., only resolver, only DoH server…) | Repeatable across setups, fast onboarding |
| DNS-Over-HTTPS stack | Out-of-the box DNS-over-HTTPS server, with a DNS-Over-HTTP server running being traefik | Allows easy setup of a DNS-over | Easy realistic lab setup for DoH paths |

Together, these features turn DoHExfTlk into a practical testbed: the same pipeline automate generation, capture, feature extraction and prediction. This makes the model comparisons fair and repeatable. In short, the toolkit is not only the foundation for Section 4, but a standalone contribution that enables researchers to conduct measurable, reproducible and explainable experiments for research and defensive engineering.

## 5.3 Limitations

### 5.3.1 Evaluation limitations

Despite its value, the evaluation remains constrained by controlled lab conditions and it's still subject to the public dataset's limitations. Benign/malicious distributions may not reflect enterprise traffic mixes or newer DoH client behaviours (e.g., HTTP/3/QUIC, cache). DoHLyzer's current flow building (short-flow cleanup and clumping windows) tend to collapse very fast bursts into one flow, reducing the number of samples available making them artificially easier to classify. On the other end stealth profiles fragment payloads across many short flows, increasing classification difficulty and exposing sensitivity to flow-construction choices. Moreover, the lack of diverse data being exfiltrated (only Appendix B has been used a file to be exfiltrated), limits variability in payload entropy, size distribution and chunking patterns. These factors may bias comparisons between each model.

### 5.3.2 Toolkit limitations

While effective as a lab testbed, the current version of DoHExfTlk toolkit has several constraints. First, it is intentionally lab-oriented rather than production-oriented: packet capture relies on Docker socket which is insecure, and host network interface card auto-discovery, which are convenient for experiments but risky in production environment. These settings create unnecessary flaws in the infrastructure, they should be replaced with least-privilege capture (e.g., without Docker socket) and read-only mount with strict RBAC (Role-Based Access Control). Also, protocol coverage is limited (no HTTP/3/QUIC) which can be used by attackers to perform DoH file exfiltration, and model packaging is sensitive to library version, meaning portability and reproducibility can break across machines.

## 6. Future work

Beyond offline evaluation, a next step is to move the toolkit from a detector to a protector by adding real-time traffic analysis and risk-based response. For example, temporarily hold-and-inspect suspicious exchanges (queue the request) or throttle the internet connection to limit exfiltration whilst the traffic is analysed at a deeper level. In parallel, the next step for the toolkit is also to handle more protocol such as HTTP/3 or QUIC, and validate its method on diverse enterprise traces, whilst generating more adverse configurations. Finally, the exfiltration client should be improved by being able to generate benign background traffic during file exfiltration (e.g., browser-like DoH lookups). This will allow the exfiltration to be camouflaged with plausible benign activity, thus giving the toolkit a more real-world like approach to scenario generation, evaluation and benchmarking of detection strategies. A next step is also to analyse when DoH file exfiltration becomes operationally uneconomical under stealth constraints. That is when, hold-and-inspect delays drive the time to exfiltrate so high that this attack method is not worth it anymore for the attacker. Going forward, the toolkit should test diverse file sizes and types to quantify how payload characteristics affect detectability, informing attackers on what files can be exfiltrated without notice while defenders can be able to know which files to protect first.

# 7. Conclusion

Objective 1 was delivered through the critical review (Section 2), which maps methods, and open problems. Objective 2 was met by deploying a toolkit that creates a testbed with features described in Table 13. Objective 3 was achieved by adapting DoHLyzer to extract flow-features as shown in Section 3.3. Objectives 4 and 5 were accomplished by training machine-learning models and integrating them with threshold-based approaches which is described in Section 3.4. Objective 6 was met via diverse evasion strategies built with the literature review in mind, as shown in Section 3.5.5. Objective 7 was covered by the Section 4, with evaluations on public datasets and the toolkit's scenarios. In the end, objective 8 was completed by releasing DoHExfTlk, a Dockerised modular and reproducible toolkit that encapsulates all the DoH-file exfiltration components and procedures.

This dissertation provides a practical and reproducible way to evaluate and compare detection strategies against malicious DNS-Over-HTTPS traffic, notably file exfiltration over DoH. To do so, it introduced DoHExfTlk, an all-in-one, containerised toolkit that automates experiments. It provides configurable exfiltration generation, to resolver-side file reconstruction, feature extraction and multi-model prediction.

Empirically, the results show that tree-based models (Gradient Boosting, Random Forest) consistently deliver near-perfect detection in controlled settings, while Logistic Regression struggles with the non-linear structure of the flows. Fixed-threshold rules (DoHxP) are fragile, they appear decisive on bursty scenarios yet having a high false positive rate on benign DoH. The analysis also highlights a flow-formation effect: very fast bursts collapse into a few highly anomalous flows which are easy to flag, whereas stealthy strategies fragment the traffic across many short flows, which impacts detection performance.

However, the work has limitations. The evaluation was conducted in a controlled lab and relies on a public dataset whose benign/malicious distributions may not match enterprise traffic or newer client behaviours.

Looking ahead, the most impactful direction this toolkit could go is to evolve from detector to protector, by adding real time feature extraction and analysis, a scoring system and risk-based response that contain exfiltration without harming user experience. It should also be able to generate benign background traffic during exfiltration to have more realistic data.

In summary, this dissertation delivers both evidence that modern ML can be effective against DoH exfiltration in controlled conditions, and a mean to test, compare, each detection model and methods. DoHExfTlk turns claims and suppositions into a measurable, reproducible experiments, providing a practical path towards deployed detection and containments methods regarding DoH file exfiltration.

# Appendices

## Appendix A: Features schema

| Column | Type | Unit | Description |
| --- | --- | --- | --- |
| SourceIP | String (IPv4/IPv6) | | Source IP address of the flow |
| DestinationIP | String (IPv4/IPv6) | | Destination IP (DoH resolver) |
| SourcePort | Integer | | Source Port |
| DestinationPort | Integer | | Destination Port (often 443) |
| TimeStamp | Float | Seconds | Flow start timestamp |
| Duration | Float | Seconds | Flow duration |
| FlowBytesSent | Integer | Bytes | Bytes sent (src -> dst) |
| FlowSentRate | Float | Bytes/second | Average send rate |
| FlowBytesReceived | Integer | Bytes | Bytes received (dst->src) |
| FlowReceivedRate | Float | Bytes/second | Average receive rate |
| PacketLengthVariance | Float | Bytes$^2$ | Variance of packet lengths |
| PacketLengthStandardDeviation | Float | Bytes | Standard deviation of packet lengths |
| PacketLengthMean | Float | Bytes | Mean packet length |

| PacketLengthMedian | Float | Bytes | Median packet length |
| --- | --- | --- | --- |
| PacketLengthMode | Float | Bytes | Modal packet length |
| PacketLengthSkewFromMedian | Float | | Skewness from median (length) |
| PacketLengthSkewFromMode | Float | | Skewness from mode (length) |
| PacketLengthCoefficientofVariation | Float | | Coefficient of variation (length) |
| PacketTimeVariance | Foat | Seconds$^2$ | Variance of inter-packet times |
| PacketTimeStandardDeviation | Float | Seconds | Standard deviation of inter-packet times |
| PacketTimeMean | Float | Seconds | Mean inter-packet time |
| PacketTimeMedian | Float | Seconds | Median inter-packet time |
| PacketTimeMode | Float | Seconds | Modal inter-packet time |
| PacketTimeSkewFromMedian | Float | | Skewness from median (time) |
| PacketTimeSkewFromMode | Float | | Skewness from mode (time) |

| PacketTimeCoefficientofVariation | Float | | Coefficient of variation (time) |
|---|---|---|---|
| ResponseTimeTimeVariance | Float | Seconds$^2$ | Variance of request ↔ response deltas |
| ResponseTimeTimeStandard | Float | Seconds | Standard deviation of request ↔response deltas |
| ResponseTimeTimeMean | Float | Seconds | Mean request ↔response deltas |
| ResponseTimeTimeMedian | Float | Seconds | Median request ↔response delta |
| ResponseTimeTimeMode | Float | Seconds | Modal request ↔ response delta |
| ResponseTimeTimeSkewFromMedian | Float | | Skewness from media (request ↔ response delta) |
| ResponseTimeTimeSkewFromMode | Float | | Skewness from mode (request ↔ response delta) |
| ResponseTimeTimeCoefficientofVariation | Float | | Coefficient of variation (request ↔ response delta) |

| Label | | String | | Benign or Malicious |
|---|---|---|---|---|

## Appendix B: File used for exfiltration.

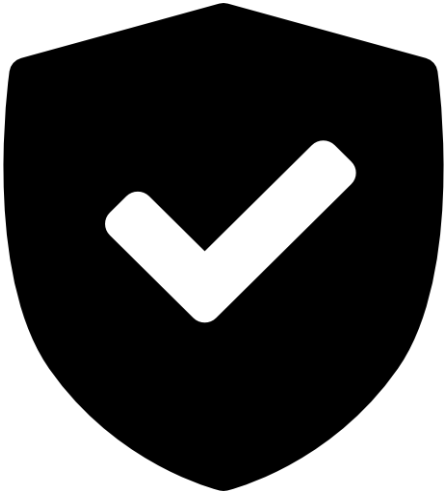

*Figure 18. File used for exfiltration*
*(https://github.com/AdamLBS/DohExfTlk/blob/main/exfiltration/client/test_data/image.png)*